\documentclass[aip,jcp,amsmath,amssymb,reprint,longbibliography]{revtex4-2}

\usepackage{graphicx}% Include figure files
\usepackage{color}
\usepackage{mathptmx}

\begin{document}

\title{Two-Dimensional Terahertz Spectroscopy of Condensed-Phase Molecular Systems}

\author{Klaus Reimann}
\email{reimann@mbi-berlin.de} 
\author{Michael Woerner}
\email{woerner@mbi-berlin.de}
\author{Thomas Elsaesser}
\email{elsasser@mbi-berlin.de}
 
\affiliation{Max-Born-Institut f\"ur Nichtlineare Optik und Kurzzeitspektroskopie, 12489 Berlin, Germany}%

\date{\today}

\begin{abstract}
Nonlinear terahertz (THz) spectroscopy relies on the interaction of matter with few-cycle THz pulses of electric field amplitudes up to
megavolts/centimeter (MV/cm). In condensed-phase molecular systems, both resonant interactions with elementary excitations at low frequency such as intra- and intermolecular vibrations and nonresonant field-driven processes are  
relevant. Two-dimensional THz (2D-THz) spectroscopy is a key method for following nonequilibrium processes and dynamics of excitations to decipher the underlying interactions and molecular couplings. This article addresses the state of the art in 2D-THz spectroscopy by discussing the main concepts and illustrating them with recent results. The latter include the response of vibrational excitations in molecular crystals up to the nonperturbative regime of light-matter interaction and field-driven ionization processes and electron transport in liquid water.
\end{abstract}

\maketitle

\section{\label{sec:intro}Introduction}

Molecular systems in the condensed phase display a variety of low-energy excitations in the frequency range from some 10 gigahertz (GHz) to 30 terahertz (THz). Collective molecular motions, inter- and intramolecular vibrations and coupled nuclear-electronic degrees of freedom such as soft modes give rise to dipole-allowed absorption and/or Raman bands, most of which exhibit complex and partly overlapping line shapes. While linear dielectric, far-infrared, and Raman spectroscopies have been developed over decades and reached a high level of sophistication,\cite{KR03B,Schrader1995} nonlinear THz methods have opened a new field of research, in which the nonlinear low-frequency response, the intrinsic dynamics of excitations, their couplings, and the interaction with a thermal bath are addressed most directly.\cite{LU04C,GA07I,HE08B,KA11A,EL19A} This rapidly developing field has benefitted from major progress in generating THz few-cycle pulses with electric field amplitudes reaching the MV/cm range \cite{BA05C,SE08B,HE08A,SA19A} and from the introduction of new nonlinear spectroscopies in which THz pulses induce a nonlinear response of a solid or a molecular ensemble.

The concept of two-dimensional correlation spectroscopy goes back to nuclear magnetic resonance \cite{JE79A} and has been transferred to the optical domain to study vibrational excitations in the mid-infrared range and/or electronic and vibronic excitations in the visible and ultraviolet. Both pump-probe and three-pulse photon-echo methods have been applied to follow the nonlinear response of a molecular system in third order in the optical field, mostly under conditions of resonant excitation. Detailed accounts of 2D third-order spectroscopy have been given in Refs.~\onlinecite{MU00B,JO03E,HA11B,CH09C}. Two-dimensional THz (2D-THz) spectroscopy, introduced some 10 years ago, \cite{KU09E,KU11B,WO13A} goes well beyond the third-order limit. It allows for mapping nonlinearities up to the regime of nonperturbative light-matter interaction, where the coupling to the external THz electric field is comparable to or even stronger than interactions within the molecular system.\cite{EL19A} Both resonant and nonresonant interactions as well as field-driven processes of charge transport have been studied by 2D-THz spectroscopy, which is inherently phase-resolved and gives access to absolute rather than relative optical phases.

Early work in 2D-THz spectroscopy has focused on method development and applications to electronic excitations in solids. Here, the large transition dipoles, e.g., of intersubband transitions in semiconductor quantum wells,\cite{KU09E,KU11A,KU11B} facilitated the generation of a quasi-resonant nonlinear response with moderate peak electric fields in the range of 1 to 50~kV/cm. More recent applications of 2D-THz spectroscopy in spectroscopic and transport studies of solids, metamaterials, and semiconductor devices have been reported in Refs.~\onlinecite{,BO14B,BO14A,SO14A,FO15B,LU17B,LU18A,KR18B,HO19A,TAR20b,WO19A,KU08A,MA17C,RA20A,MA21A}. The rapid development of generation schemes for THz pulses with MV/cm amplitudes has strongly widened the application range of 2D-THz methods, now including vibrational and/or phononic excitations and, in particular, a broad range of nonresonant nonlinear interactions. This article gives an account of the current state of this field with a focus on molecular excitations and processes in the condensed phase. It combines a presentation of the method and experimental aspects with a discussion of recent prototypical applications.

The content of this article  is organized as follows. Section~\ref{sec:THzpulses} summarizes methods for THz pulse generation in laser-driven sources (Sec.~\ref{sec:generation}) and field-resolved detection (Sec.~\ref{sec:field}), including a short description of linear THz spectroscopy. Section \ref{sec:2dTHz} introduces the basic concepts of 2D-THz spectroscopy and the experimental implementation (Sec.~\ref{sec:concept}), complemented by a discussion of data handling and analysis (Sec.~\ref{sec:analysis}). A comparison to related 2D methods such as nonlinear 2D Raman-THz and multicolor 2D spectroscopy is included as well (Sec.~\ref{sec:rel2d}). A prototypical study of soft-mode excitations in molecular crystals of aspirin is presented in Sec.~\ref{sec:softmode}, followed by a discussion of ionization and electron transport in water driven by strong THz fields (Sec.~\ref{sec:water}). Conclusions and an outlook on future developments are given in Sec.~\ref{sec:conc}.

\section{\label{sec:THzpulses}Few-cycle TH\lowercase{z} pulses: generation and characterization }

\subsection{\label{sec:generation}Laser-driven generation of TH\lowercase{z} pulses}

\begin{figure}
	\includegraphics[width=\columnwidth]{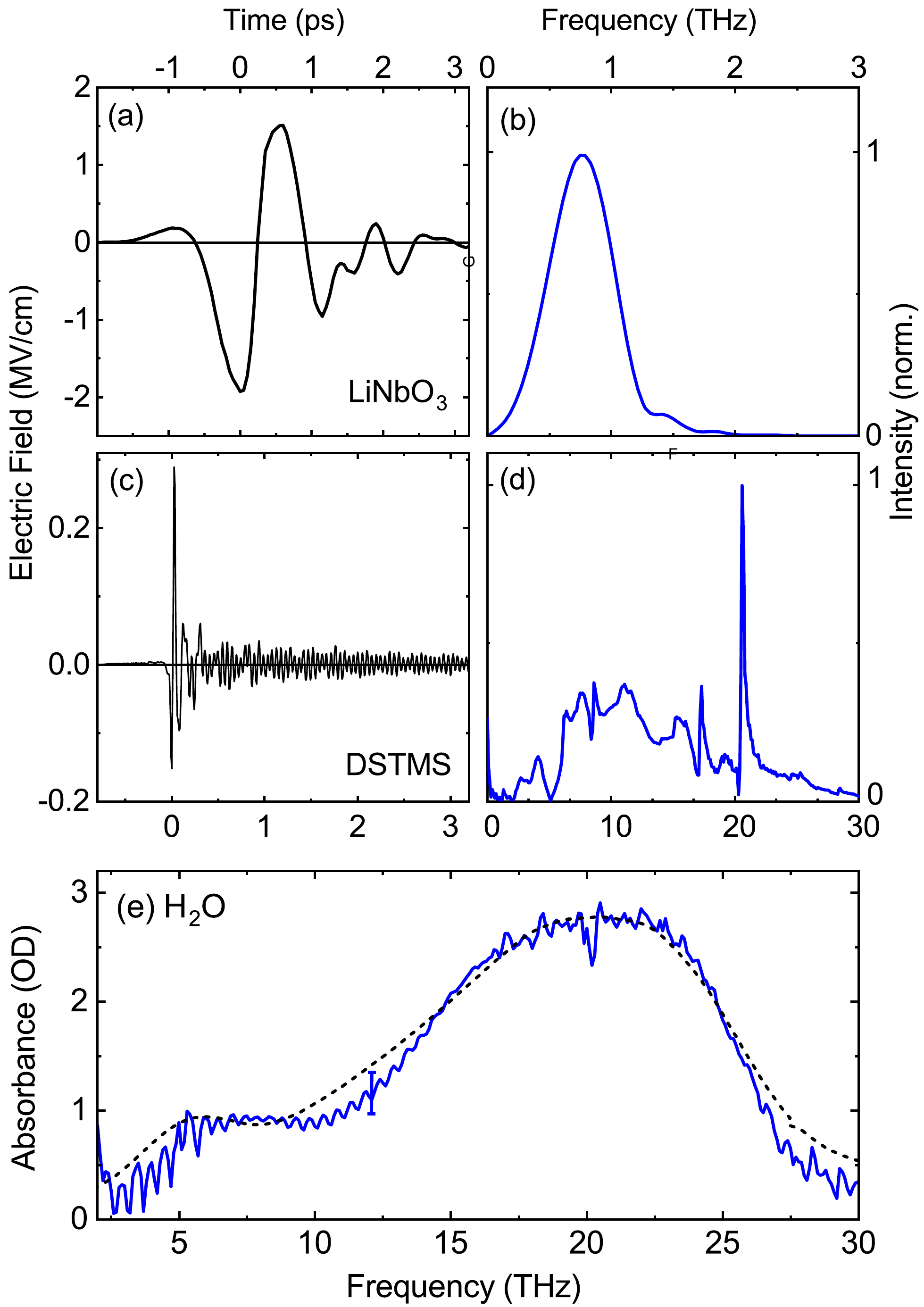}
	\caption{\label{fig:epsart} \label{fig:epsart} THz pulses generated by optical rectification of femtosecond optical pulses in (a,b) LiNbO$_3$ and (c,d) the organic crystal DSTMS. Panels (a) and (c) show the electric field as a function of time, panels (b,d) the intensity spectra of the pulses. Please note the different frequency scales in (b) and (d). (e) THz absorption spectrum of a 15~$\mu$m thick water sample with a high peak absorbance of 2.8 (blue line, taken from Ref. \onlinecite{FO17D}). The absorbance $A=-\log(T)$ ($T$: intensity transmission of the sample) derived from a time-domain measurement with the broadband pulses shown in panels (c) and (d) is plotted as a function of frequency. The dashed line gives a reference spectrum from literature. \cite{BE96C}}\label{fig:gen}
\end{figure}

Nonlinear THz spectroscopy relies on ultrashort THz pulses with an electric field amplitude sufficient to induce a nonlinear response in the sample under study. Such pulses have been generated with electrically biased photoconductive switches and antenna structures or by nonlinear optical frequency conversion of optical pulses from the visible and near-infrared to the THz range. Irradiation of a photoconductive switch made from GaAs, InP or other semiconductors by a femto- to picosecond optical pulse induces  short-lived transient currents, which radiate electric fields at THz frequencies. The maximum amplitude of the THz field is set by the DC bias field of the switch and reaches values on the order of 10~kV/cm for large-area emitters. While photoconductive switches driven by femtosecond laser oscillators are widely applied in commercial devices for linear THz spectroscopy, their applicability in nonlinear experiments is limited because of the comparably small peak electric fields.

Plasmas generated with intense femtosecond laser pulses in dilute gases emit strong few-cyle THz pulses.\cite{BA05C,CO00B,KR04A} Two-color generation of a plasma in air or nitrogen, e.g. by the fundamental and second-harmonic frequency of an amplified femtosecond laser pulse, generates a unidirectional transient plasma current of electrons, which radiates a very broad frequency spectrum  including pronounced THz components (up to 70~THz).\cite{BA10B}  Peak electric fields of up to 400~kV/cm have been generated at a kilohertz repetition rate with 25-fs pulses from an amplified Ti:sapphire laser system.\cite{BA05C} This type of source has been applied in a number of nonlinear THz experiments, although plasma instabilities result in fluctuations of the electric field amplitudes and the spatial beam parameters.

Optical difference frequency mixing or rectification in nonlinear crystals and/or at surfaces with a high second-order nonlinearity $\chi^{(2)}$ is the predominant method for generating few-cycle THz pulses with high electric field amplitudes. Here, the difference frequency of two synchronized incoming pulses or between different components in the broad spectrum of a single femtosecond pulse is generated, preferentially in a phase-matched conversion process, in which the phase velocity of the THz pulse needs to match the group velocity of the pump pulses. Beyond the THz frequency range discussed here, this method has been widely applied to generate pulses in the mid-infrared up to frequencies on the order of 100~THz.

Inorganic semiconductors with a zincblende crystal structure such as GaAs, GaP, InP, CdTe, and ZnTe and the ferroelectric LiNbO$_3$ are standard materials to cover a frequency range up to some 6~THz with electric field amplitudes\cite{EL19A} up to several hundred kV/cm.  The introduction of non-collinear rectification in LiNbO$_3$ using pump pulses with a tilted wavefront \cite{HE08A,BOD19a} allows for reaching peak electric fields of several MV/cm at frequencies between 0.2 and 1.5~THz. The time-dependent electric field of a THz pulse generated with this method in shown in Fig.~\ref{fig:gen}(a), the corresponding intensity spectrum in Fig.~\ref{fig:gen}(b). Drawbacks of this method are the spatial and temporal break-up of the optical pump due to angular dispersion, which limits the conversion efficiency, and the limited quality of the generated THz beam. Very recent work has addressed such issues both by numerical simulations of different interaction geometries and experiments. \cite{WA20A} It should be noted that few-cycle pulses with electric field amplitudes up to 100~MV/cm have been generated\cite{SE08B} in the frequency range from 10 to 70~THz.

Organic nonlinear crystals consisting of molecular chromophores have received substantial interest for THz generation because of their high second-order nonlinearities. \cite{ZH07B,VI15B} Figures \ref{fig:gen}(c) and (d) display the time-resolved electric field and the intensity spectrum of pulses generated by driving a 400-$\mu$m thick DSTMS crystal (DSTMS: 4-N,N-dimethyl\-amino-4'-N'-methyl\-stilbazolium 2,4,6-tri\-methyl\-benzene\-sul\-fo\-nate) with 25-fs pulses from an amplified Ti:sapphire laser.\cite{SO15D} The THz spectrum covers the range from 0.1 to 30~THz. The narrow features in the spectrum originate from radiation on vibrational resonances and correspond to the long-lived oscillations of the electric field in the time-domain. The frequency conversion process is only partially phase-matched. However, due to the very high optical nonlinearity, an appreciable conversion efficiency is possible even without phase-matching. The comparably low optical damage thresholds of organic nonlinear crystals at higher pulse repetition rates eventually limit the attainable electric field amplitudes.

\subsection{\label{sec:field}Field-resolved detection and measurement}

Phase-resolved detection of THz pulses, i.e., the measurement of the electric field as a function of time, is a key ingredient of 2D-THz spectroscopy. The standard method is free-space electrooptic (EO) sampling, which exploits the change of refractive index induced by the THz field in an electrooptic crystal.\cite{WU95B} This change is mapped by the polarization change of a femtosecond probe pulse, the duration of which needs to be much smaller than the THz period.  Measuring the polarization change for different delay times between the THz transient and the probe pulse allows for reconstructing the time-dependent THz field and, in calibrated setups, the absolute electric field amplitudes. Widely used electrooptic materials include ZnTe, GaP, GaSe, LiTaO$_3$, and LiNbO$_3$. A detection bandwidth on the order of 100~THz has been demonstrated with 10 to 50~$\mu$m thick ZnTe and GaSe crystals. \cite{HU00A,KU04B} The smallest electric fields measured by EO sampling are on the order of 1~V/cm, as shown in studies of vacuum fluctuations of the electric-field. \cite{RI15A,BE19A} Further details and literature can be found in Ref.~\onlinecite{EL19A}. A setup for EO sampling is presented in Sec.~\ref{sec:concept} (Fig.~\ref{fig:3dsetup}).

Time-domain THz methods have found broad application in linear mid- and far-infrared spectroscopy. A field-resolved detection scheme allows for deriving both the real and imaginary part of the linear dielectric function from the THz field transmitted through a sample. Moreover, the fact that electric fields are detected rather than intensities as in conventional far-infrared spectroscopy makes the study of optically thick samples possible. The electric field amplitude decreases only proportional to the square root of the intensity, e.g., a reduction of intensity to 1/100 corresponds to a reduction of 1/10 in electric field amplitude only. This aspect is relevant for aqueous solutions, e.g., of biomolecules where the background absorption of water represents a major contribution to the overall absorption spectrum. To illustrate the potential of field-resolved spectroscopy, Fig.~\ref{fig:gen}(e) shows the absorption spectrum of a 15~$\mu$m thick film of neat water (blue line), derived from a time-domain THz measurement\cite{FO17D} with the broadband pulses shown in Figs.~\ref{fig:gen}(c) and (d). The spectrum, which is in good agreement with literature data (dashed line, Ref.~\onlinecite{BE96C}), displays the prominent L2 absorption band around 20~THz (600~cm$^{-1}$) due to librations of water molecules, and the weaker stretching band of OH$\cdots$O hydrogen bonds between 5 and 6~THz. The peak absorbance in this spectrum has a value of approximately 2.8, corresponding to a very small intensity transmission of $1.5 \times 10^{-3}$, which is typically beyond the detection range of conventional Fourier spectrometers.

\section{\label{sec:2dTHz}Two-dimensional TH\lowercase{z} spectroscopy}

\subsection{\label{sec:concept}Concept and experimental implementation}

\begin{figure}
	\includegraphics[width=\columnwidth]{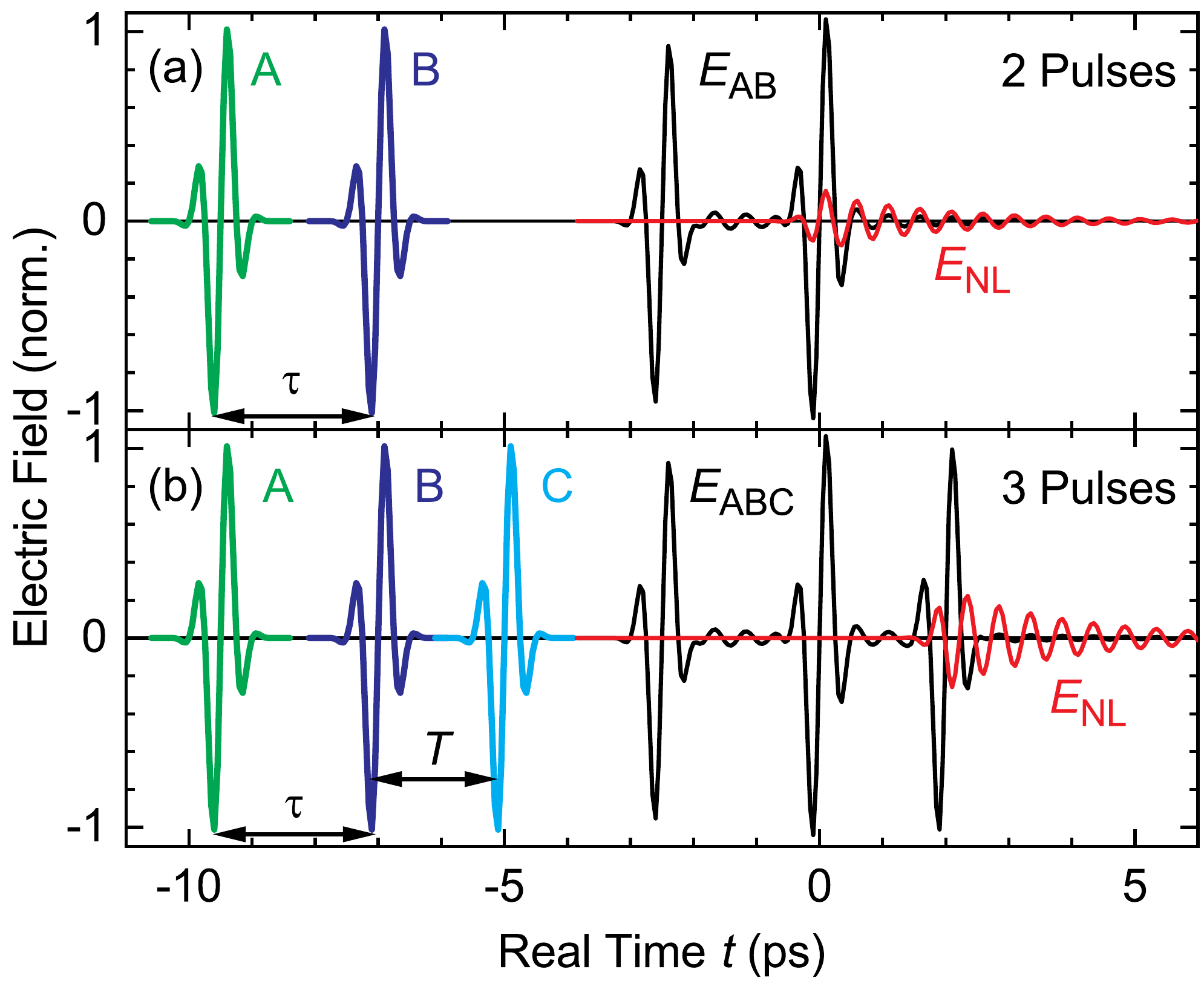}
	\caption{Calculated electric field transients of (a) a two-pulse and (b) a three-pulse sequence. The transients labeled A, B, C are the incident pulses before interaction with the sample. $E_\text{AB}$ and $E_\text{ABC}$ are the transients after transmission through the sample when two [in (a)] or three [in (b)] pulses have interacted with the sample. $E_\text{NL}$ is the nonlinear signal according to Eqs.~\eqref{eq:2dNL} and \eqref{eq:3dNL}. The nonlinear response was calculated for an ensemble of two-level systems. \label{fig:pulses}}	
\end{figure}

Two-dimensional THz spectroscopy is based on the interaction of a pulse sequence consisting of two or three THz pulses with a sample and a measurement of the total transmitted THz electric field in amplitude and phase. If the response of the sample is perfectly linear, the total transmitted field $E_\text{ABC}^\text{lin}$ of, e.g., three pulses (A, B, C) is equal to the sum of the electric fields of the individual pulses:
\begin{equation}
	E^\text{lin}_\text{ABC}=E^\text{lin}_\text{A}+E^\text{lin}_\text{B}+E^\text{lin}_\text{C}\label{eq:linresp}.
\end{equation} 
Accordingly, 2D-THz spectroscopy only yields additional information in case of a nonlinear response of the sample. In 2D-THz spectroscopy, one exploits the nonlinearities resulting from interaction with all pulses applied. For two pulses [Fig.~\ref{fig:pulses}(a)], the nonlinear response determines the electric field $E_\text{NL}(\tau,t)$ given by \cite{KU09E}
\begin{equation}
	E_\text{NL}(\tau,t) = E_\text{AB}(\tau,t)-E_\text{A}(\tau,t)-E_\text{B}(t).\label{eq:2dNL}
\end{equation} 
In this equation, $\tau$ is the time delay between the two pulses A and B, $t$ the real time, i.e., the time axis along which the electric field is measured by electrooptic sampling, $E_\text{A}(\tau,t)$ the transmitted  electric field when only pulse A is incident on the sample, $E_\text{B}(t)$ the transmitted field when only pulse B is incident on the sample, and $E_\text{AB}(\tau,t)$ the transmitted field with both pulses present. The single-pulse nonlinearities are contained in $E_\text{A}(\tau,t)$ and $E_\text{B}(t)$, so that they are removed by Eq.~\eqref{eq:2dNL}. 

For three pulses [Fig.~\ref{fig:pulses}(b)], one has to subtract both one- and two-pulse nonlinearities to recover $E_\text{NL}(\tau,T,t)$ according to \cite{SO16C}
\begin{eqnarray}
	E_\text{NL}(\tau,T,t) &=& E_\text{ABC}(\tau,T,t)\nonumber \\
	&&- E_\text{AB}(\tau,T,t)-E_\text{BC}(T,t)-E_\text{AC}(\tau,T,t)\nonumber\\
	&&+E_\text{A}(\tau,T,t)+E_\text{B}(T,t)+E_\text{C}(t).\label{eq:3dNL}
\end{eqnarray}   
The delay between pulses B and C is $T$, often called waiting time. The meaning of $E_\text{ABC}(\tau,T,t)$ etc. is analogous to the two-pulse case. 
For two pulses, the  nonlinear susceptibilities contributing to $E_\text{NL}$ are $\chi^{(2)}$, $\chi^{(3)}$, and higher, for three pulses they are $\chi^{(3)}$, $\chi^{(4)}$, and higher. In general, an $n$-pulse scheme of this type will show nonlinearities of order $n$ and higher. 

Nonlinearities of even and odd orders can be distinguished by looking at the spectrum of the nonlinear field. For input fields of a frequency $\omega$ even-order nonlinearities will result in nonlinear fields with frequencies 0, $2\omega$, $4\omega$, $\ldots$, whereas odd-order nonlinearities generate fields with $\omega$, $3\omega$, $5\omega$, $\ldots$. Thus, a nonlinear signal with the frequency $\omega$ in a two-pulse experiments can not result from a second-order nonlinearity but originates from a third- or higher odd-order nonlinearity. Most third-order nonlinearities can be measured with only two pulses. 

The different contributions to the total nonlinear signal can be analyzed and separated with density matrix theory describing the nonlinear light-matter interaction.\cite{MU95A} The signal components correspond to different pathways in Liouville space. Such pathways have been visualized by diagrammatic techniques such as double-sided Feynman diagrams, \cite{MU95A} multi-level schemes, \cite{PO92B} or sequences of arrows in 2D frequency space.\cite{WO13A}

\begin{figure}
	\includegraphics[width=\columnwidth]{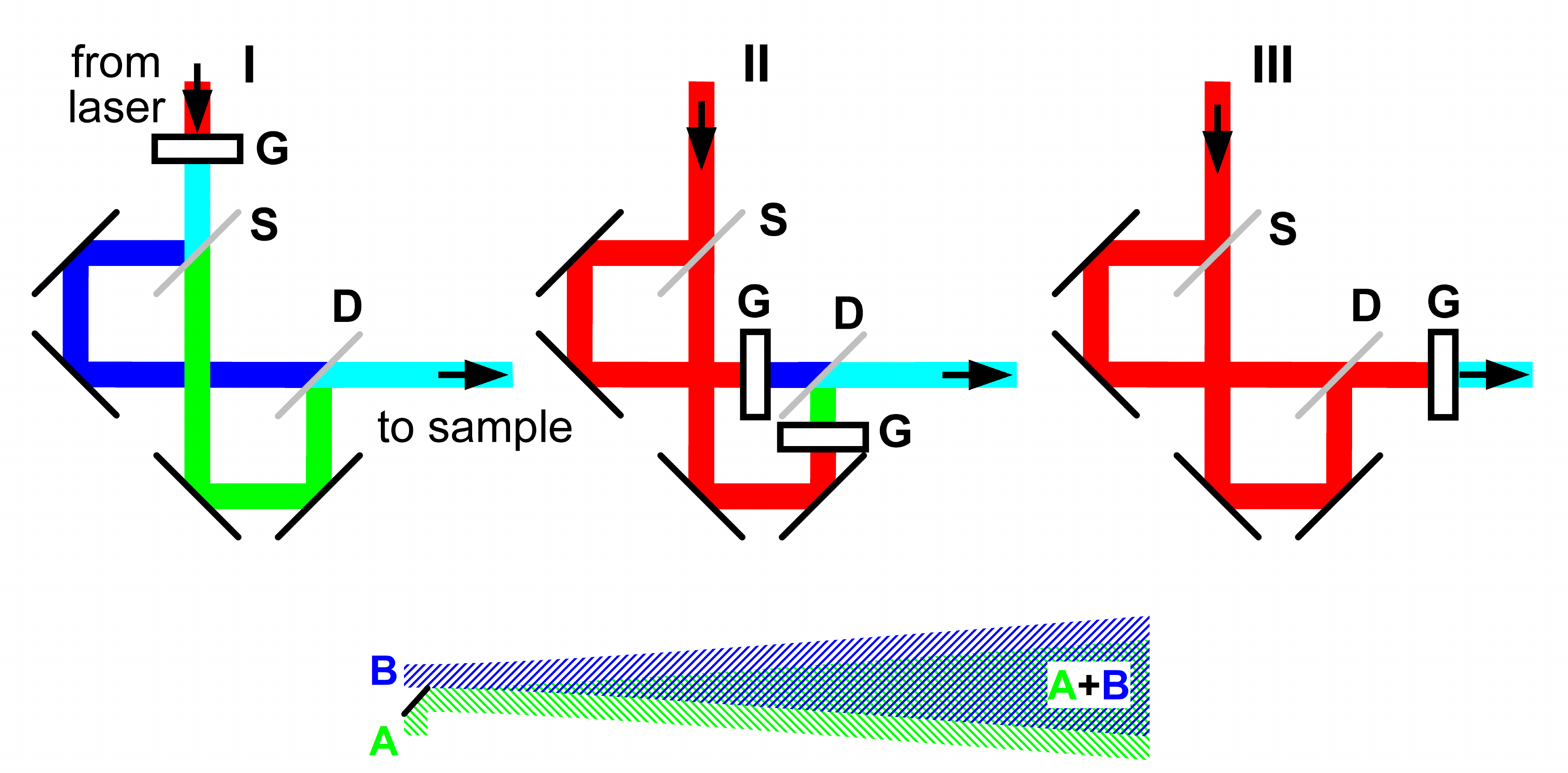}
	\caption{Upper part: Schemes for generating a sequence of two THz pulses from a visible/near-infrared laser. G consists of a nonlinear crystal for THz generation including a filter to remove the residual laser light, S and D are beam splitters. The laser beam is shown in red, the two THz beams in blue and green.\label{fig:split} Bottom part: Overlapping divergent THz beams A and B as an alternative to using a beam splitter D in schemes I and II. \label{fig:diffrac}}	
\end{figure}
 
Reproducible femtosecond delays between the THz pulses and synchronization with the read-out pulse for electrooptic sampling are essential for 2D-THz spectroscopy. Typically, all pulses are derived from a single mode-locked laser oscillator.  There are various schemes to generate the two or three THz pulses for 2D-THz spectroscopy as schematically illustrated in Fig.~\ref{fig:split} for two pulses. In scheme I, a single THz pulse is generated, split in two replicas, and recombined after variable time delays. In scheme II, the beam from the pump laser is split and then sent to two separate THz generators G. In this way it is possible to have different frequencies of the two beams for two-color 2D spectroscopy. Scheme III requires only a single THz generator G, but has the disadvantage that the generation process leads to nonlinear electric fields according to the definitions \eqref{eq:2dNL} and \eqref{eq:3dNL}, in particular during the time overlap of the pump pulses. 

Key components of schemes I and II are the THz beam splitters S and D in I and D in II. An ideal beam splitter should generate a reflected pulse with $E_r(t) =a\,E(t-\delta_r)$ and a transmitted pulse $E_t(t) =b\,E(t-\delta_t)$ from the incident THz pulse $E(t)$ with $a^2+b^2=1$. Here, $\delta_t$ and $\delta_r$ are time delays upon reflection and transmission. Components coming close to this ideal are thin pellicle beam splitters, plate-type beam splitters, and wire grid polarizers.  The thin plastic substrate of a pellicle introduces weak dispersion-induced changes of the transmitted transients, but may display considerable absorption at certain THz frequencies. Disadvantages of pellicles are their fragility and their high sensitivity to acoustic noise and air currents. Thicker plate-type beam splitters need to possess low THz dispersion and absorption. Suitable materials are undoped diamond, silicon, and germanium. Unwanted reflections from the back surface of beamsplitter plates can be suppressed by antireflection coatings.\cite{TH08A,BRA14c,CAI17c,CHA15v} Another solution is to use the beam splitter under Brewster's angle for $p$ polarization and add a metallic coating on one side for the required reflectivity. \cite{SO17B} 
Wire grid polarizers consist of parallel thin metallic wires. For wavelengths large compared to the distance between neighboring wires, they transmit electric fields perpendicular to the direction of the wires and reflect electric fields parallel to the wires. After a wire grid beamsplitter D, the two beams will be orthogonally polarized. To obtain the same polarization, one can introduce an additional polarizer between D and the sample.

\begin{figure}
	\includegraphics[width=\columnwidth]{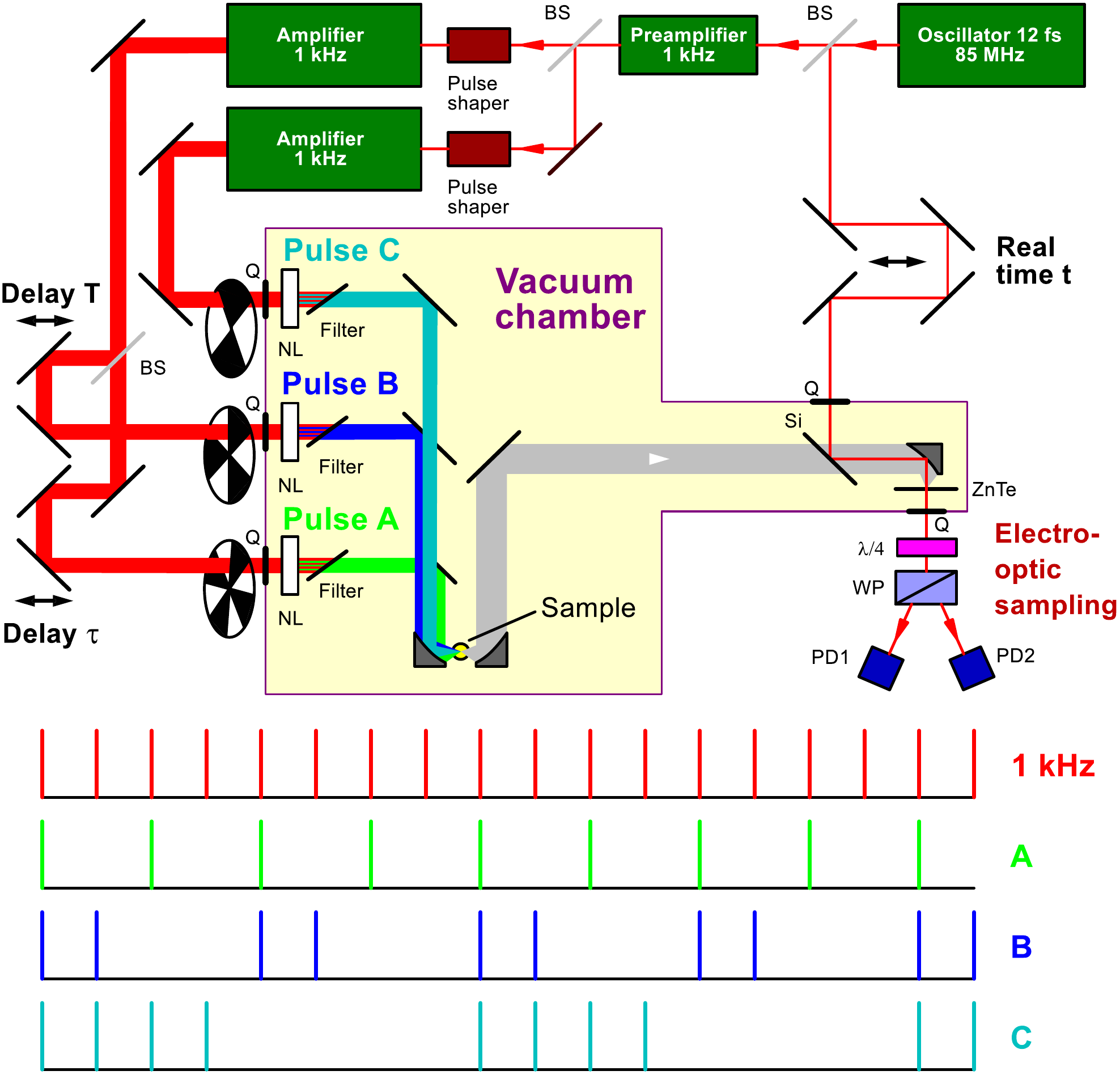}
	\caption{Setup for three-pulse measurements. BS are beam splitters, Q are quartz glass windows for the vacuum chamber, NL are nonlinear crystals for THz generation (GaSe), ZnTe is the electrooptic crystal, $\lambda/4$ is a quarter-wave plate, WP is a Wollaston polarizer, and PD1 and PD2 are silicon photodiodes. The detected signal is the difference of the outputs of PD1 and PD2. Details are discussed in the text.\label{fig:3dsetup}}	
\end{figure}

In both schemes I and II, the beam splitters D can be replaced by having the two beams parallel and close to each other (lower part of Fig.~\ref{fig:diffrac}). Because of the generally high divergence of THz beams, they will overlap after some propagation length. For instance,  an initial beam waist of 5~mm (typical aperture of nonlinear crystals, e.g., GaSe) at a frequency of 1~THz (wavelength $\lambda=300~\mu$m) results in a minimum divergence of 38~mrad. After a propagation length of 30~cm the two beams overlap nearly perfectly. This method can easily be extended to three beams arranged, e.g., in a triangular pattern (Fig.~\ref{fig:3dsetup}). In such geometries, essentially no losses occur, i.e., in scheme I for two (three) beams each beam has, upon hitting the sample, one half (one third) the energy of the initially generated THz pulse outside the time overlap of the two pulses. In contrast, the pulse energies on the sample is only one quarter (one ninth) of the initial energy for ideal beam splitters D.  

A setup for three pulses using scheme II is shown in Fig.~\ref{fig:3dsetup}.\cite{SO16C} The pump pulses are generated with a Ti:sapphire oscillator-amplifier system. This particular setup includes two multipass amplifiers, both seeded from the same oscillator to ensure synchronization. In this scheme, the shape of the two amplified pulses can be set independently with acousto-optic pulse shapers, permitting both phase and amplitude changes. The settings are chosen for optimum generation of the required THz pulses,\cite{LU04B} a feature particularly important for two-color measurements.  

The setup of Fig.~\ref{fig:3dsetup} uses pulses from the oscillator as read-out pulses in electrooptic sampling. They are shorter than the amplified pulses (12~fs versus 20-25 fs) and, thus, allow for measuring THz pulses at higher frequency (40~THz vs. 25~THz).\cite{RE07C,EL19A} Moreover, the smaller pulse-to-pulse fluctuations improve the signal-to-noise ratio. While the temporal jitter between oscillator and amplifier output is typically on the order of a few femtoseconds, temporal drifts of the read-out relative to the THz pulse represent a challenge in this scheme. Due to the long optical path length through the amplifiers (on the order of 10 m), even a very small temperature difference between the path of the oscillator pulse and the path of the amplified pulse can lead to significant drifts. For example, a temperature difference of 0.1~K between parts of the optical table will lead to a change of the distance of 10~$\mu$m, equivalent to a time difference of 33~fs. It is, however, possible to correct for such drifts after the measurement (see Sec.~\ref{sec:analysis}). 

The read-out and the THz pulse overlap in a thin (thickness $\approx 10~\mu$m) electrooptic ZnTe crystal of (110) orientation. The thin crystal is mounted onto a 0.5~mm-thick (100) ZnTe crystal, which shows no electrooptic effect. \cite{LE99E} In this way, the thin electroptic crystal is mechanically more stable, and, even more important, the time range over which reflections from the back surface of the crystal are absent, is extended. The THz electric field changes the dielectric tensor of the (110) ZnTe crystal, which in turns leads to a change of the polarization state of the initially linearly polarized read-out pulse. This change is converted by the quarter-wave plate and the polarizer into a difference between the signals of two balanced photodiodes. The difference signal is proportional to the THz electric field at this instant of time. It is digitized for every laser shot and stored in a PC, which also detects the settings of all choppers (see below) and moves the various delays according to a predetermined sequence. 

The THz generation stages, the sample, and the EO sampling setup are placed in a closed chamber, which is evacuated with a scroll and a turbomolecular pump for removing gases such as CO$_2$ and water vapor. In this way, absorption and temporal distortions of the THz pulses are suppressed. At a pressure of $10^{-3}$~mbar, the gas concentrations are reduced from their values at atmospheric pressure by six orders of magnitude.
At pressures of the order of $10^{-6}$~mbar it is possible to perform measurements at low temperatures by mounting the sample on the cold finger of a helium-flow cryostat. 
%The sample is precisely aligned with manual manipulators and one can switch between samples while they are cold.

Mechanical choppers synchronized to the 1-kHz repetition rate of the Ti:sapphire amplifiers are introduced in all beams to distinguish the THz pulses in the pulse sequence. The chopper in beam A works with half the repetition rate (500~Hz), B with one quarter (250~Hz), and C with one eighth (125~Hz). In this way chopper A transmits one pulse and blocks one, B transmits two and blocks two, and C transmits four and blocks four (lower part of Fig.~\ref{fig:3dsetup}). In this way, all possible combinations of the three pulses A, B, and C are measured. Varying the delays $T$, $\tau$, and $t$ one obtains the single-pulse transients $E_\text{A}(T,\tau,t)$, $E_\text{B}(T,t)$, and $E_\text{C}(t)$, the two-pulse transients $E_\text{AB}(T,\tau,t)$, $E_\text{BC}(T,t)$, and $E_\text{AC}(\tau,t)$, the three-pulse transient $E_\text{ABC}(T,\tau,t)$, and the nonlinear signal [Eq.~\eqref{eq:3dNL}].

\begin{figure*}
	\includegraphics[width=\textwidth]{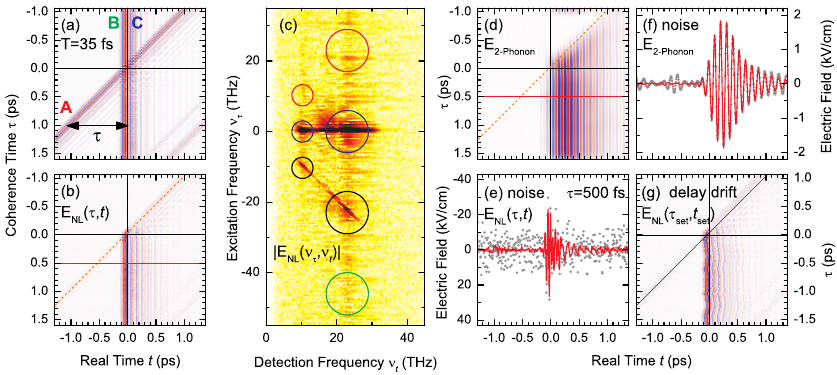}
	\caption{Two-dimensional spectroscopy on InSb using three THz pulses as shown in Fig.~\ref{fig:pulses}(b). (a)~Contour plot of the sum of electric field transients $E_\text{A}(\tau,T,t)+E_\text{B}(T,t)+E_\text{C}(t)$ transmitted through the InSb sample as a function of the coherence time $\tau$ and real time $t$ for the waiting time $T=35$~fs. (b)~Nonlinear signal $E_{\text{NL}}(\tau,t)$ for $T=35$~fs according to Eq.~\eqref{eq:3dNL}. The orange dashed line indicates the center of pulse A. The linear amplitude scales of the 2D scans range over (a)~$\pm 76.5$~kV/cm and (b)~$\pm 30$~kV/cm. (c)~Contour plot of the amplitude $|E_{\text{NL}}(\nu_{\tau},\nu_t)|$ which is the 2D Fourier transform of $E_{\text{NL}}(\tau,t)$. The colored circles indicate the position of relevant signals in the 2D frequency space. (d)~Selecting the peak at the two-TO-phonon resonance (small blue circle at $\nu_{\tau}=0,\nu_t=10$~THz) by a 2D-gaussian filter allows for back-transforming the $E_{\text{2-Phonon}}(\tau,t)$ signal to the time domain. Red transients in (e) and (f) show $E_{\text{NL}}(\tau=500~\text{fs},t)$ and $E_{\text{2-Phonon}}(\tau=500~\text{fs},t)$, respectively. Grey circles: Signals after artificially adding white noise to the measured $E_{\text{NL}}(\tau,t)$ shown in panel (b). (g)~$E_{\text{NL}}(\tau_\text{set},t_\text{set})$ as a function of the set-point values $\tau_\text{set}$ and $t_\text{set}$ given by the computer controlling the experiment. Part of the results shown here have been presented in Refs.~\onlinecite{SO16C,SO16D}.}\label{fig:InSb}
\end{figure*}

\subsection{\label{sec:analysis}Data analysis}

Two-dimensional THz spectra are derived from the electric field $E_\text{NL}(\tau,T,t)$ at a fixed waiting time $T$ by a 2D Fourier transform along $\tau$ and $t$, giving the signal $E_\text{NL}(\nu_\tau,\nu_t)$ as a function of the excitation frequency $\nu_\tau$ and the detection frequency $\nu_t$. For illustrating this concept and discussing key aspects of data analysis, we use a 2D-THz data set recorded in the three-pulse setup of Fig.~\ref{fig:3dsetup} with the narrow-gap semiconductor InSb.\cite{SO16C,SO16D} 

InSb is a direct-gap III-V semiconductor with a band gap of $E_{g}=0.17$~eV,  corresponding to a frequency of $\nu= E_{g}/h = 41$~THz.\cite{KA57A} There is a TO two-phonon resonance at 10 THz, as observed in the higher-order Raman spectrum. The THz pulses with a center frequency of 21~THz are neither resonant to the band gap nor to the two-phonon resonance. The THz peak field strength $E\approx50$~kV/cm and the extraordinarily large interband transition dipole $d_\text{cv}=e\cdot(4$~nm$)$ at the $\Gamma$ point ($e$: elementary charge)\cite{KA57A,KA59A}  result in a Rabi frequency $\Omega_\text{Rabi}=E\,d_\text{cv}/\hbar=30$~THz, i.e., comparable to the THz carrier frequency. As a result, the 2D-THz experiments are in the strongly nonperturbative regime of light-matter interaction, which is
characterized by the simultaneous occurrence of nonlinear polarizations of different orders $\chi^{(n)}$. In particular, the nonperturbative regime allows for multiple interactions of a single THz pulse with the InSb crystal. In Refs.~\onlinecite{SO16C,SO16D}, we reported the ultrafast dynamics of two-phonon coherences and signals related to multiple two-photon interband excitation of electron-hole pairs. In the following, we concentrate on the dynamics of two-TO-phonon coherences in InSb.

In Fig.~\ref{fig:InSb}(a), the sum of electric field transients $E_\text{A}(\tau,T,t)+E_\text{B}(T,t)+E_\text{C}(t)$ transmitted through the InSb sample is shown in a contour plot as a function of the coherence time $\tau$ and real time $t$ for the waiting time $T=35$~fs. The corresponding nonlinear signal $E_{\text{NL}}(\tau,T,t)$ derived from Eq.~\eqref{eq:3dNL} is plotted in panel (b). In accordance with causality, a nonvanishing $E_{\text{NL}}(\tau,T,t)$ sets in only with or after the last pulse in the timing sequence. For better orientation, the center of pulse A is indicated by the orange dashed line, which intersects the horizontal $\tau=0$ line (black) at $t=-T$. In Fig.~\ref{fig:InSb}(c), the contour plot of the amplitude $|E_{\text{NL}}(\nu_{\tau},\nu_t)|$ is displayed for $T=35$ fs.

The circles of different size and color indicate the positions of relevant signals in the 2D frequency space. As a function of detection frequency $\nu_t$, strong nonlinear signals occur in the spectral range of the driving pulses $15\;\text{THz}<\nu_t<25\;\text{THz}$ (large circles) and at the two-phonon resonance $\nu_t=10$~THz (small circles). The former signals have been discussed in detail in Ref.~\onlinecite{SO16D}. In the following we focus on the small circles at the two-phonon resonance $\nu_t=10$~THz. As a function of excitation frequency $\nu_{\tau}$ we observe significant nonlinear signals for $\nu_{\tau}=0$ (blue circles), $\nu_{\tau}=-\nu_t$ (black circles), and $\nu_{\tau}=+\nu_t$ (red circles). The signal at $(\nu_\tau,\nu_t)=(0,10)$~THz is a nonlinear signal of at least 11th order in the THz field, as has been discussed in detail in Ref.~\onlinecite{SO16C}. For $\tau > 0$, i.e., for the pulse sequence (A$(\tau,T,t)$, B$(T,t)$, C$(t)$) pulse A induces a complete two-photon absorption event, i.e., it creates an electron-hole pair by four interactions with its electric field. After the coherence time $\tau$, pulse B creates a two-phonon coherence via an impulsive third-order Raman excitation, which evolves during the waiting time $T$. Pulse C generates a second electron-hole pair, again requiring four interactions and thereby transferring the two-phonon coherence to the final electronic state. After the third pulse, the two-phonon coherence evolves along the real time $t$ and emits radiation via its optical transition dipole.

Selection of the peak at $(\nu_\tau,\nu_t)=(0,10)$~THz by a 2D-Gaussian filter and subsequent Fourier back-transform into the time-domain gives the signal field $E_{\text{2-phonon}}(\tau,t)$,  which is plotted in Fig.~\ref{fig:InSb}(d) as a function of $\tau$ and $t$. Panel (f) shows a cut of this signal field for $\tau=500$~fs. The Fourier filtering by the 2D Gaussian results in an extraordinarily low noise level below $<0.1$~kV/cm. For comparison, a cut of the total signal field $E_\text{NL}(\tau,t)$ for $\tau=500$~fs and $T=35$ fs is shown in Fig.~\ref{fig:InSb}(e). Here, a noise level of $\approx 2$~kV/cm is estimated from the signal at $t<0$, which should vanish for causality reasons. This analysis shows that 2D-Fourier filtering can considerably increase the signal-to-noise ratio of narrow signal peaks in the 2D-THz spectrum. To further illustrate the power of 2D-Fourier filtering, we added artificially white noise to the measured $E_{\text{NL}}(\tau, t)$ shown in panel (b), which results in the grey dots in panels (e) and (f). Here, the total nonlinear signal $E_{\text{NL}}(\tau=500~\text{fs}, t)$ is barely detectable. In contrast, the 2D-filtered nonlinear two-phonon coherence $E_{\text{2-phonon}}(\tau=500~\text{fs}, t)$ has a noise level of $\approx 0.3$~kV/cm [grey dots for $t<0$ in panel (f)] allowing for a clear experimental characterization of this nonlinear contribution.     

We now address the correction of long-term delay drifts between the three THz pulses A, B, and C derived from the output of the Ti:sapphire amplifiers in Fig.~\ref{fig:3dsetup} and the read-out pulse from EO sampling which is
derived from the mode-locked oscillator. In Fig.~\ref{fig:InSb}(g), the nonlinear signal field $E_\text{NL}(\tau_\text{set},t_\text{set})$ is plotted as a function of the uncorrected times $\tau_\text{set}$ and $t_\text{set}$, 
as given by the computer controlling the delay stages. The time drifts in this raw signal lead to slight distortions of the shape and the period of the signal wavefronts. Such distortions can easily be corrected as the timing of the individual three pulses is measured simultaneously during a complete 2D scan, applying the pulse sequences shown in the lower part of Fig.~\ref{fig:3dsetup}. Using a 2D linear interpolation, one can reconstruct the nonlinear signal on a regular equidistant $(\tau,t)$ 2D grid resulting in the 2D signal $E_{\text{NL}}(\tau,t)$ shown in Fig.~\ref{fig:InSb}(b).

The separation of nonlinear signals with respect to their order in the THz field and the assigment to different pathways in Liouville space (cf. Sec. \ref{sec:concept}) requires distinct non-overlapping signal peaks in the frequency-domain 2D spectrum. In the nonperturbative regime of both resonant and nonresonant light-matter coupling, this condition may not be fulfilled. Resonantly excited Rabi flops, e.g., of intersubband transitions, \cite{KU09E} include signal components up to a very high nonlinear order and display overlapping 2D signal peaks at high driving fields. Electrons coherently driven in the conduction band of a semiconductor by a strong nonresonant THz field display a nonperturbative response and emit a multitude of harmonics of the fundamental driving frequency which are difficult to separate.\cite{KU10D,SC14A} In general, an interplay of resonant and nonresonant contributions to the overall nonlinear response results in highly complex 2D-THz spectra. Here, variation of the phase and polarization of the THz pulses and, in anisotropic samples, measurements with different sample orientations can help to decipher the 2D spectra.

\subsection{\label{sec:rel2d}Related 2D Methods}

There is a number of related 2D methods which involve at least one THz pulse in a sequence of ultrashort pulses and/or rely on the detection of coherent THz emission from an excited sample. In the present context,  
2D Raman-THz, \cite{SA13B,HA17B} 2D THz-Raman,\cite{AL15C,FIN16a,JO19A} and THz-infrared-visible spectroscopy \cite{GR18A} are relevant, all  so far being applied in the third-order limit of light-matter interaction.
       
Two-dimensional Raman-THz spectroscopy of liquids gives insight in the dynamics and couplings of intermolecular degrees of freedom. The method involves a femtosecond pulse in the visible or near-infrared for the Raman interaction, which is second order in the electric field, and a THz pulse which induces a coherent polarization by a single interaction with the transition dipole of a low-frequency excitation. The THz emission of the sample represents the third-order nonlinear signal, which is measured as a function of real time $t_2$ in amplitude and phase by EO sampling. The second relevant time axis is the temporal separation $t_1$ of the Raman and the THz pulse. 

Two interaction sequences have mainly been applied in experiments on liquid water and aqueous salt solutions.\cite{SA13B,HA17B,HA12B,SHA17g,CI19A,SI19A} For positive $t_1$, The Raman pulse interacts with the sample first and generates a Raman coherence, which is translated to a dipole-induced coherence by an interaction with the delayed THz pulse. The latter coherence gives rise to THz dipole emission, the nonlinear signal. For negative $t_1$, the THz pulse comes first and generates a dipole coherence,  which is switched to a Raman coherence by two interactions with the delayed Raman pulse. The THz signal field is emitted at a time $t_2$ after the THz pulse.

Two-dimensional Raman-THz spectroscopy of water and aqueous salt solutions has revealed echo-like signals in the range of water-water hydrogen-bond modes. A detailed account on such results has been given in Ref.~\onlinecite{HA17B}. The ultrafast structural fluctuations of aqueous systems  set a time window, in which a particular local structure exists and rephasing of the vibrational excitations of different environments is possible. As a result, 2D Raman-THz spectroscopy represents a direct probe of the complex intermolecular dynamics in liquids. A correct theoretical description of the nonlinear Raman-THz response requires polarizable water models, which can be benchmarked by comparison with 2D spectra.\cite{HAM14a,IK15B,PA15A,ITO16c}  

Two-dimensional THz-Raman spectroscopy implies two interactions with THz pulses and one interaction with the near-infrared Raman pulse. \cite{FIN16a,FI17B} In a three-level system, interaction with the first THz pulse generates a coherence on the 1-2 transition, which the second THz pulse at a delay $t_1$ transfers to the 2-3 transition. This coherent excitation is read out after a time interval $t_2$ by a near-infrared Raman pulse on the 3-1 transition.
The coherences during the periods $t_1$ and $t_2$ allow for generating 2D spectra as a function of $t_1$ and $t_2$ or---after a 2D Fourier transform---as a function of the corresponding frequencies $\nu_1$ and $\nu_2$.  Anharmonic couplings of low-frequency modes in liquid CHBr$_3$, CCl$_4$, and CBr$_2$Cl$_2$ have been studied by 2D THz-Raman spectroscopy.\cite{AL15C,FIN16a,FI17B} In part, these results have been re-analyzed in Ref.~\onlinecite{ME20A}.  

Third-order THz-infrared-visible spectroscopy \cite{GR18A} aims at elucidating couplings between low- and high-frequency vibrational modes and involves a THz pulse, a femtosecond mid-infrared pulse, and a sub-picosecond near-infrared or visible pulse. The THz and the mid-infrared pulse are both in resonance with vibrational transitions, so that each generates a vibrational coherence. The third interaction with the visible pulse transfers the coherent polarization to the visible spectral range and, thus, generates a coherent emission at the frequencies ($\nu_\text{VIS}+\nu_\text{IR}\pm\nu_\text{THz}$). There is a delay $t$ between the THz pulse and the mid-infrared and visible pulses, the latter with a fixed temporal delay to each other.  The emitted electric field is heterodyned with a local oscillator for sensitive phase-resolved detection. From this signal, 2D spectra are calculated as a function of $\nu_1$, which is in the THz range and derived from a Fourier transform along $t$, and as a function of the mid-infrared frequency $\nu_2$, which is obtained by subtracting $\nu_\text{VIS}$ from the frequency of the nonlinear signal. Details of data processing and an analysis have been given in Ref.~\onlinecite{VBG19}.  THz-infrared-visible spectroscopy has been applied to phonons in solids and to liquid water.   

We note that there are a number of 2D experiments in which a coherent THz emission has been recorded after interaction of the sample with pulses at higher frequencies. A recent example is a study of TO phonon coherences in bulk GaAs. \cite{GH20B}

\begin{figure*}
	\includegraphics[width=0.95\textwidth]{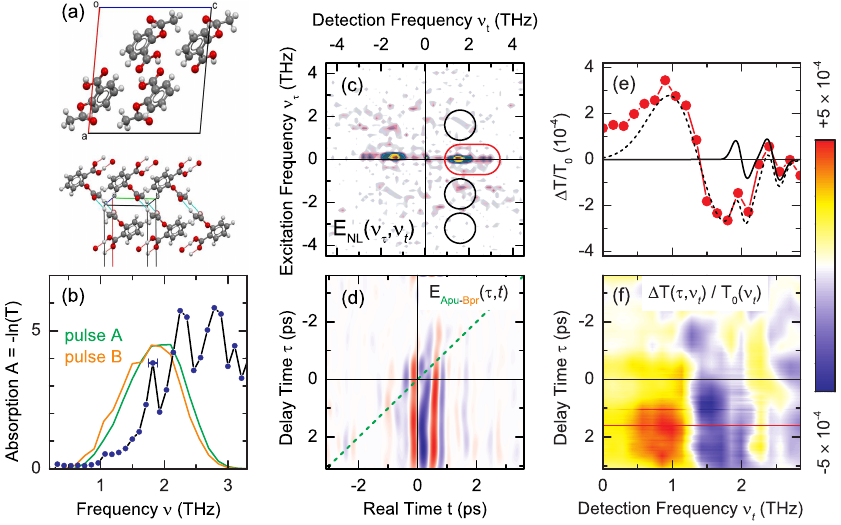}
	\caption{(a)~Unit cell and crystal structure of aspirin. (b)~Linear absorption of a polycrystalline aspirin pellet measured at a temperature of 80~K together with the amplitude spectra of THz pulses A (green) and B (orange). The weak feature around 1.1~THz represents the soft-mode absorption. (c)~Contour plot of the nonlinear 2D spectrum $|E_\text{NL}(\nu_\tau,\nu_t)|$. The colored ovals indicate the positions $(\nu_\tau,\nu_t)$ of relevant signals in the 2D frequency space. The red oval encircles the A$_\text{pump}$--B$_\text{probe}$ signal at $(\nu_\tau=0,\nu_t)$. (d) Contour plot of $E_\text{Apu-Bpr}(\tau,t)$ (amplitude scale: $\pm 150$~V/cm) from a back transform of the 2D Fourier-filtered spectrum. The green dashed line indicates the center of pump pulse A. (e)~Spectrally resolved A$_\text{pump}$--B$_\text{probe}$ signal (symbols) at a delay time of $\tau =1.6$~ps [indicated by the red horizontal line in panel (f)]. The dashed line gives the calculated field-induced response dominated by the softmode, the solid line the minor contributions of the modes at 1.8 and 2.3 THz. (f)~Contour plot of the spectrally resolved A$_\text{pump}$--B$_\text{probe}$ signal as function of the probe frequency $\nu_t$ and pump-probe delay $\tau$. The results shown here have been presented in Ref.~\onlinecite{FO17C}.}\label{fig:Aspirin}
\end{figure*}

\section{\label{sec:softmode}Soft-mode excitations in polar molecular crystals}

Soft modes are lattice vibrations with a particularly strong coupling to the electronic system of a polar and/or ionic crystal. Excitation of a soft mode is connected with a pronounced relocation of electronic charge, thus directly affecting the electric polarization of the material. Vice versa, the coupling to electronic degrees of freedom enhances the vibrational oscillator strength significantly. Prototypical systems displaying soft modes are displacive ferroelectrics, e.g., with a perovskite lattice structure, and other ionic crystals. The interplay of soft-mode excitations and changes of electronic charge density has been revealed in femtosecond x-ray diffraction experiments following the transient charge densities in space and time.\cite{ZA12A,HA19A} Much less is known on the intrinsic nonlinear response of soft modes,\cite{KA12B} in particular in molecular materials. In the following, we discuss results from 2D-THz spectroscopy which give detailed insight in soft mode nonlinearities in aspirin crystals.\cite{FO17C}
       
The crystal structure of aspirin consists of layers, in which pairs of aspirin molecules are arranged as cyclic hydrogen-bonded dimers [Fig.~\ref{fig:Aspirin}(a)]. The layers are connected via hydrogen bonds between the acetyl groups of neighboring molecules. Theoretical calculations of the electronic and vibrational structure of aspirin  have shown that a number of vibrational modes in the THz range display a significant coupling to the electronic system.\cite{RE14A} Among them is the rotation of the methyl groups at a frequency of 1.1 THz, much lower than that of a free CH$_3$ rotator.
Figure~\ref{fig:Aspirin}(b) shows the linear THz absorption of a polycrystalline aspirin sample at a temperature of 80~K as measured with weak THz pulses. The methyl rotation gives rise to the weak absorption shoulder around 1~THz. A 2D-THz experiment was performed with two pulses A and B, the spectra of which are shown in Fig.~\ref{fig:Aspirin}(b). The maximum electric field amplitudes of pulses A and B are 25 and 50~kV/cm.

The 2D-THz spectrum shown in Fig.~\ref{fig:Aspirin}(c) exhibits different pump-probe and photon-echo signals, among which the A$_\text{pump}$-B$_\text{probe}$ signal at $\nu_\tau=0$ is the strongest one (oval boundary). Applying a 2D Fourier filter (see Sec.~\ref{sec:analysis}) and transforming the signal back to the time domain gives the contour plot of Fig.~\ref{fig:Aspirin}(d), which shows an oscillation along $t$ with a frequency below 2~THz, the maximum of the pulse spectra in Fig.~\ref{fig:Aspirin}(b). Moreover, there is a phase shift compared to the field of pulse B (not shown). To assess this transient in more detail, it was Fourier-transformed along the real time $t$ to generate the spectrally resolved pump-probe signal as a function of $\nu_t$ and the delay time $\tau$ between the two pulses [Fig.~\ref{fig:Aspirin}(e) and (f)]. The symbols in panel (e) give the pump-probe signal at a delay time $\tau=1.6$~ps, which shows a bleaching (increased transmission) for frequencies $\nu_t < 1.4$~THz and an enhanced absorption (decreased transmission) at higher frequencies. This behavior persists up to delay times of several picoseconds [Fig.~\ref{fig:Aspirin}(f)]. 

Pulse A in the 2D experiment induces both a population change of the v=0 and v=1 states of the vibrations within the pump spectrum and a nonlinear polarization. The population changes result in a third-order ($\chi^{(3)}$) nonlinear response with a transmission increase on the  v=0 to 1 transition and a transmission decrease  on the anharmonically redshifted $v=1$ to 2 transition. The spectral envelope of the measured pump-probe signal in Figs.~\ref{fig:Aspirin}(e) and (f) with a \emph{blueshifted} transmission decrease demonstrates that such population-induced signals play a minor role here. Instead, the signal is dominated by a pump-induced blueshift of the $v=0$ to 1 transition of the soft mode and represents a response beyond the $\chi^{(3)}$ regime. 

The observed behavior reflects the transient local-field response of the aspirin crystal. Due to the strong enhancement of the soft-mode transition dipole by coupling to the electronic system, there is a strong dipole-dipole coupling between different aspirin sites, a concomitant electric polarization and a strong Lorentz electric field. In thermal equilibrium, the electrostatic energy of the crystal is minimized, resulting in a pronounced redshift of the soft-mode frequency compared to its value without a Lorentz field. The transition frequency of 1.1 THz observed in linear THz absorption spectrum is a result of this redshift. Excitation by pulse A in the 2D-THz experiment reduces the contribution of the excited molecules to the overall polarization. This nonlinear saturation of electric polarization reduces the local electric field and shifts the soft mode back to a higher frequency, i.e., a blueshift arises. This nonperturbative mechanism governs the line shapes of the pump-probe signals in Figs.~\ref{fig:Aspirin}(e) and (f) and strongly dominates over the population-induced signals. This interpretation is supported by an analysis of the photon-echo signals in the 2D-THz spectrum [Fig.~\ref{fig:Aspirin}(c)], which has been presented in Ref.~\onlinecite{FO17C}. The photon-echo signals display pronounced components at negative delay times $\tau$, which originate from non-instantaneous contributions of Lorentz fields with the time structure of a free-induction decay.  

The 2D-THz results reveal a nonlinear optical response of the aspirin soft mode in the nonperturbative regime, even for moderate THz driving fields of some 50 kV/cm. This fact reflects the sensitivity of collective electric properties, here the local field, to a comparably weak external stimulus. A similar behavior is expected for a large range of vibrational and/or phononic excitations with a pronounced coupling to the electronic system. The interplay of vibrational excitations and electronic properties is highly relevant for optically induced phase transitions of crystal structure and for transiently steering macroscopic electric properties of functional materials. 

\begin{figure}
	\includegraphics[width=\columnwidth]{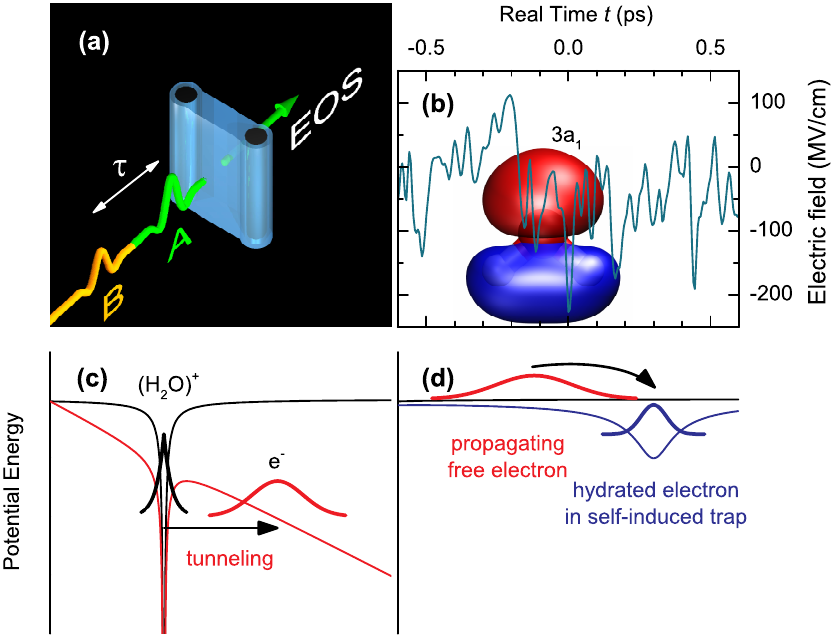}
	\caption{(a)~Schematic of the 2D-THz experiment with phase-locked THz pulses A and B spearated by delay $\tau$, the water jet (blue) in transmission geometry, and the path toward the electrooptic sampling (EOS) detector. (b)~Fluctuating electric fields in liquid water as calculated from a molecular dynamics simulation for electrons at the position of the second-highest occupied	orbital $3a_1$ (HOMO$-$1). (c)~Schematic diagram of
	field-induced electron tunneling. (d) Schematic of electron propagation	and localization.}\label{fig:H2Ointro}
\end{figure}

\section{\label{sec:water}Field-induced ionization and electron transport in liquid water}

\begin{figure*}
	\includegraphics[width=0.95\textwidth]{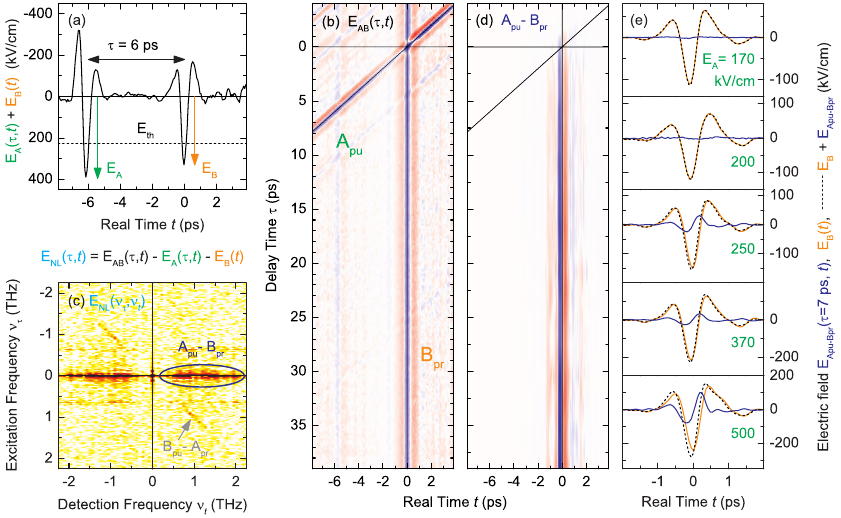}
	\caption{Time-resolved nonlinear THz response of liquid water. (a)~THz pulses $E_\text{A}(\tau=6\;\text{ps},t)$ and $E_\text{B}(t)$ transmitted through a $50\;\mu$m-thick liquid water jet. (b)~Contour plot of the 2D-THz scan $E_\text{AB}(\tau,t)$ transmitted through the water sample as a function of real time $t$ and delay time $\tau$ between pulses A and	B. (c)~2D-Fourier transform $E_\text{NL}(\nu_\tau,\nu_t)$ of $E_\text{NL}(\tau,t)$ as a function of detection frequency $\nu_t$ and excitation frequency $\nu_\tau$. The 2D spectrum shows a predominant A$_\text{pump}$--B$_\text{probe}$ signal (blue ellipse) and a much weaker B$_\text{pump}$--A$_\text{probe}$ signal (grey arrow). d)~Contour plot of the A$_\text{pump}$--B$_\text{probe}$ signal gained by Gaussian 2D-Fourier filtering and back transform to the 2D time domain. (e)~Cuts of time-resolved A$_\text{pump}$--B$_\text{probe}$ signal (blue lines) at a fixed delay time $\tau=7$ ps for different peak amplitudes E$_A$ of pulse A between 170 to 500~kV/cm. The probe field E$_\text{B} (t)$ is shown as an orange line.
		The sum $E_\text{B}+E_\text{NL}$ (dashed lines) shows that electron generation leads to both a phase shift to earlier times $t$ and an amplitude increase of the probe pulse B.}    
	\label{fig:H2Odata}
\end{figure*}

Intermolecular and/or collective molecular motions together with librations, i.e., hindered rotations, give rise to the complex absorption spectrum of liquid water in the frequency range from a few GHz to some 30 THz [Fig.~\ref{fig:gen}(e)]. The broad unstructured envelopes and the substantial spectral overlap of different absorption bands make an analysis of linear spectra in terms of the underlying degrees of freedoms highly demanding. This fact has generated a quest for nonlinear time-resolved spectroscopies, by which the different excitations can be discerned in a better way and couplings between modes be identified. 

A first proposal for nonlinear 2D-THz spectroscopy of liquid water has been published in the 1990s. \cite{OK98A} Recent work in 2D Raman-THz and THz-infrared-visible spectroscopy has started to address the properties of intermolecular vibrations (see Sec.~\ref{sec:rel2d}). There has been related work studying the nonlinear Kerr effect in a similar frequency range.\cite{PA96D,WI00C,HA05C,KA18A,KA18B,TC19A} Until very recently, however, there have been no genuine 2D-THz results on liquid water, partly due to the insufficient experimental sensitivity and comparably small transition dipole moment of intermolecular and collective modes. In the following, we discuss very recent results of a 2D-THz study of liquid water, which applies THz pulses with peak electric field amplitudes in a range from 100 kV/cm to 2 MV/cm to study the nonperturbative response of the liquid.\cite{GH20A} 

The experimental geometry is sketched in Fig.~\ref{fig:H2Ointro}(a). A phase-locked pair of pulses A and B separated by a delay $\tau$ interacts with a 50-$\mu$m thick water jet. As the water jet cannot be operated in vacuum, it was placed in an atmosphere of dry nitrogen, together with the 2D-THz setup. The transmitted pulses are recorded by EO sampling (EOS). A transmitted pulse pair detected this way is shown in Fig.~\ref{fig:H2Odata}(a). The pulses cover a spectral range from 0.2 to 2~THz. The nonlinear signal E$_\text{NL}(\tau,t)$ is given by Eq.~\eqref{eq:2dNL}. In Fig.~\ref{fig:H2Odata}(b), the field E$_\text{AB}(\tau,t)$ is plotted as a function of $\tau$ and $t$, while panel (d) shows the nonlinear signal field E$_{NL}(\tau,t)$. In the 2D frequency-domain [Fig.~\ref{fig:H2Odata}(c)], the strong pump-probe signal generated with the stronger pump pulse A and the weaker probe pulse B (A-B signal) is complemented by the much weaker B-A pump-probe signal. This observation points to a highly nonlinear dependence of $E_\text{NL}$ on the pump field, which has been studied in a series of measurements with pump pulses A of different peak field strengths $E_A$. Data for different values of $E_A$ and a pump-probe delay of $\tau=7$ ps are summarized in Fig.~\ref{fig:H2Odata}(e). Here, the orange lines show the transmitted probe pulse B measured without excitation and the blue lines the nonlinear signal field.

The results of Fig.~\ref{fig:H2Odata}(e) demonstrate a threshold behavior of the nonlinear pump-probe signal. For field amplitudes of pulse A below the threshold $E_\text{th}= 225$~kV/cm, the nonlinear signal (blue lines) vanishes for all delay times $\tau$. Above the threshold, E$_\text{NL}(\tau=7$ps,$t)$ shows a dispersive shape. It reaches considerable amplitudes up to 100~kV/cm. As seen from the total transmitted probe field (dashed lines), the nonlinear response leads to an increase in amplitude and a phase shift along the $t$ axis, i.e., there is a nonlinear change of both the real and imaginary part of the refractive index of the excited water jet. A quantitative analysis discussed in detail in Ref.~\onlinecite{GH20A} gives a decrease of the real and imaginary part by some 10 \%, with a flat spectral envelope distinctly different from the absorption spectrum shown in Fig.~\ref{fig:gen}(e). 
     
The strong changes of the real and imaginary parts of the refractive index together with their spectra demonstrate that excitations of intermolecular vibrations giving rise to the comparably weak absorption bands of Fig.~\ref{fig:gen}(e) (molar  extinction coefficient $\epsilon(1~\text{THz})=1.67~\text{M}^{-1}\text{cm}^{-1}$) play a minor role for the observed nonlinear response. The same holds for the nonlinear Kerr effect in the THz range which is connected with changes of the refractive index several orders of magnitude smaller than found here. \cite{GH20A,ZA18A,EL20A} Instead, the nonlinear response is governed by the generation of solvated electrons in the presence of the THz field, strongly modifiying both the real and imaginary part of the refractive index in a wider frequency range.

Water molecules in the liquid phase have an electric dipole moment $d \approx 2.9$~Debye, which gives rise to a very strong electric field in the hydrogen bond network of H$_2$O. Due to thermally activated molecular motions at ambient temperature, such fields display fluctuations in a time range from approximately 50~fs to several picoseconds. An electric field trajectory from molecular dynamics simulations is shown in Fig.~\ref{fig:H2Ointro}(b), where the electric field projected on the $3a_1$ electron orbital of a water molecule is plotted as a function of time. Individual fluctuation peaks reach amplitudes on the order of 100 to 200 MV/cm.

At such high external fields, the binding potential of electrons in the water molecule is strongly distorted and tunneling ionization becomes possible, as schematically shown in Fig.~\ref{fig:H2Ointro}(c). 
As both amplitude and direction of the fluctuating electric field change on an ultrafast time scale, recombination processes of a released electron with its H$_2$O$^+$ parent ion at the ionization site strongly dominate over successful ionization events, resulting in a very low concentration of released electrons on the order of some 10$^{-7}$ M. This scenario changes dramatically if one superimposes a directed THz field on the fluctuating electric field. Now, the THz field can drive the released electron away from the ionization site and, thus, induce a persistent charge separation. After charge separation and transport, electrons are equilibrating in their own solvation shell on a time scale of a few picoseconds [Fig.~\ref{fig:H2Ointro}(d)].\cite{Herbert2019} The presence of such solvated electrons changes the dielectric function or refractive index in the THz range and, thus, causes the observed strong nonlinear response. This response is in the regime of nonperturbative light-matter interaction under nonresonant conditions. 

Independent evidence for the generation of solvated electrons comes from the observation of the characteristic absorption band of solvated electrons around 700~nm after interaction of the water sample with the strong THz field.\cite{GH20A} The strength of this absorption with a known molar extinction coefficient \cite{HA08D} allows for estimating the electron concentration c$_e$. One derives c$_e = 5 \times 10^{-6}$~M for a THz peak field of 500~kV/cm and c$_e = 2 \times 10^{-5}$~M for 1.9~MV/cm.   

The results in Fig.~\ref{fig:H2Odata}(e) demonstrate a threshold electric field $E_\text{th}= 225$~kV/cm of pulse A for the generation of solvated electrons. This threshold originates from the requirement of a persistent spatial separation of the electron from its parent ion by the transport process along the THz field. The electron needs to take up a ponderomotive (or kinetic) energy exceeding the ionization energy of a water molecule of approximately 11~eV. This energy is provided by the local THz field in the liquid which is, due to the polarity of the water molecules, roughly two times higher than the external field. Thus, an external field of some 250~kV/cm corresponds to a local field of 500~kV/cm at which the ponderomotive energy reaches a value of 11~eV, the water ionization energy. Below this threshold, recombination of the electron with its parent ion prevails, while for even higher fields the electron either travels larger distances in the liquid or generates secondary electrons by impact ionization of water molecules.

The present 2D-THz study of water in the nonperturbative regime reveals a novel aspect of the strong fluctuating electric fields in polar liquids, namely their potential for inducing tunneling ionization processes. In combination with strong THz fields, mobile electrons can be generated and their transport and localization processes be steered by tailoring the THz fields in amplitude and time. Such an approach may allow for studying a new regime of charge transport in liquids.

\section{\label{sec:conc}Conclusions}

Two-dimensional THz spectroscopy has undergone a rapid development and represents an important addition to the portfolio of multidimensional spectroscopies. It allows for studying resonant excitations and field-driven processes up to arbitrary nonlinear orders, i.e., clearly exceeds the third-order limit. In particular, the response of a molecular ensemble under conditions of a nonperturbative light-matter interaction becomes accessible.

There is a number of directions for future developments. First, a more systematic application of 2D-THz spectroscopy and related 2D methods to intermolecular modes in liquids and low-frequency phonons in crystalline and disordered solids appears promising. Secondly, the investigation of charge transport processes in molecular systems driven by nonresonant THz fields holds potential for elucidating frictional forces and the interplay between delocalized and localized states of electrons, protons, and ions. Third, field-induced changes of vibrational transition frequencies, i.e., a vibrational Stark effect induced by strong THz fields, will allow for mapping electric fields in liquids and biomolecules at their intrinsic THz fluctuation frequencies and, thus, go beyond steady-state electrostatics. Work along those lines will benefit from the ongoing rapid development of THz sources and detection systems, and the implementation of more complex pulse sequences in multidimensional THz spectroscopy.

\section*{Acknowledgments}
We thank our colleagues who were involved in the research reviewed here, in particular Klaus Biermann, Benjamin P. Fingerhut, Giulia Folpini, Christos Flytzanis, Ahmed Ghalgaoui, and Carmine Somma. We acknowledge funding by the Deutsche Forschungsgemeinschaft (WO 558/14-1) and by the European Research
Council (ERC) under the European Unions Horizon
2020 Research and Innovation Program (grant agreement no. 833365). 

\section*{Data availability}
The data that support the findings of this study are available from the corresponding author upon reasonable request.

\section*{references}
%\bibliography{literat2,THz4}
%aipnum4-2.bst 2019-01-14 (MD) hand-edited version of apsrev4-1.bst
%Control: key (0)
%Control: author (8) initials jnrlst
%Control: editor formatted (1) identically to author
%Control: production of article title (0) allowed
%Control: page (1) range
%Control: year (1) truncated
%Control: production of eprint (0) enabled
\makeatletter\ifx\inprint\@undefined\def\inprint{in print}\fi
\ifx\submitted\@undefined\def\submitted{submitted to }\fi
\ifx\unpublished\@undefined\def\unpublished{unpublished}\fi

\end{document}